\documentclass[10pt,twocolumn]{IEEEtran}
\usepackage{cite}
\usepackage{algorithmic}
\usepackage{algorithm}
\usepackage{array}
\usepackage{mdwmath}
\usepackage{mdwtab}
\usepackage{eqparbox}
\usepackage{fixltx2e}
\usepackage{url}
\usepackage{amssymb}
\usepackage[cmex10]{amsmath}
\usepackage[mathscr]{eucal}
\usepackage{bm}
\usepackage{ifpdf}
\usepackage{stmaryrd}
\usepackage{multicol}
\usepackage{stfloats}
\usepackage{xcolor}
\usepackage{subfigure}
\usepackage[dvips]{graphicx}

\newcommand{\mb}[1]{{\boldsymbol #1}}
\newcommand{\y}{\mb{y}}
\newcommand{\Y}{\mb{Y}}
\newcommand{\n}{\mb{n}}
\newcommand{\N}{\mb{N}}
\newcommand{\A}{\mb{A}}
\newcommand{\I}{\mb{I}}
\renewcommand{\S}{\mb{S}}

\newcommand{\s}{\mb{s}}
\renewcommand{\a}{\mb{a}}
\newcommand{\ar}[1]{{\a_{r}\left(\theta_{#1}\right)}}
\newcommand{\at}[1]{{\a_{t}\left(\theta_{#1}\right)}}
\newcommand{\bgv}[1]{{\left[ \begin{matrix} #1 \end{matrix} \right]}}
\newcommand{\E}[1]{\mathbb{E}\left\{#1\right\}}
\renewcommand{\Re}[1]{\mathrm{Re}\left\{#1\right\}}

\newcommand{\tr}[1]{{\mathrm{tr}}\left(#1\right)}
\renewcommand{\vec}[1]{{\mathrm{vec}}\left(#1\right)}
\newcommand{\rank}[1]{{\mathrm{rank}}\left(#1\right)}
\newcommand{\diag}[1]{{\mathrm{diag}}\left(#1\right)}
\newcommand{\B}{\mb{B}}
\newcommand{\W}{\mb{W}}
\newcommand{\J}{\mb{J}}
\newcommand{\C}{\mathbb{C}}

\newcommand{\p}{\mb{p}}
\newcommand{\Q}{\mb{Q}}
\newcommand{\G}{\mb{G}}
\newcommand{\V}{\mb{V}}
\newcommand{\F}{\mb{F}}
\newcommand{\Z}{\mb{Z}}
\newcommand{\q}{\mb{q}}
\newcommand{\g}{\mb{g}}
\newcommand{\e}{\mb{e}}
\newcommand{\f}{\mb{f}}
\renewcommand{\d}{\mb{d}}

\newcommand{\w}{\mb{w}}
\newcommand{\h}{\mb{h}}

\newcommand{\T}{\mb{T}}
\newcommand{\U}{\mb{U}}
\newcommand{\D}{\mb{D}}

\renewcommand{\H}{\mb{H}}
\newcommand{\X}{\mb{X}}
\newcommand{\obj}{\mathcal{X}}
\def \SNR {\mathrm{SNR}}
\def \INR {\mathrm{INR}}
\def\equ #1{\begin{equation}#1\end{equation}}

\newcommand{\beq}{\begin{equation}}
\newcommand{\eeq}{\end{equation}}

\newtheorem{lemma}{Lemma}

\begin{document}

\title{Robust Design of Transmit Waveform and Receive Filter For Colocated MIMO Radar}
\author{Wei~Zhu
        and Jun~Tang \thanks{The authors are with the Department of Electronic Engineering,
Tsinghua University, Beijing, 100084, China (email: zhuwei11@mails.tsinghua.edu.cn, tangj\_ee@mail.tsinghua.edu.cn).}}

\maketitle
\IEEEpeerreviewmaketitle

\begin{abstract}
We consider the problem of angle-robust joint transmit waveform and receive filter design for colocated Multiple-Input Multiple-Output (MIMO) radar, in the presence of signal-dependent interferences. The design problem is cast as a max-min optimization problem to maximize the worst-case output signal-to-interference-plus-noise-ratio (SINR) with respect to the unknown angle of the target of interest. Based on rank-one relaxation and semi-definite programming (SDP) representation of a nonnegative trigonometric polynomial,
a cyclic optimization algorithm is proposed to tackle this problem.
The effectiveness of the proposed method is illustrated via numerical examples.

\end{abstract}


\section{Introduction}
Due to many advantages over conventional phased-array radar \cite{li2009,Li07,Fishler06,Tabrikian06},
multiple-input-multiple-output (MIMO) radar has been widely studied over the last decade.
For both colocated \cite{Li07} and distributed MIMO radar \cite{Haimovich08}, one of the most crucial problems is how to design probing signals properly. Existing design approaches can mainly be classified into five categories according to the criteria adopted: 1) optimizing the radar ambiguity function \cite{Antonio07,Chen08}; 2) matching a desired beam-pattern \cite{Stoica07,Fuhrmann08,Soltanalian2014,Vorobyov2014}; 3) optimizing the detection or estimation performance based on information theory \cite{Yang07,Leshem07,tangBo2010,Nehorai2010}; 4) optimizing an estimation-oriented lower bound (e.g., Cram\'{e}r-Rao bound \cite{Li08} and Reuven-Messer bound \cite{Tabrikian13}) and 5) joint transmit waveform and receive filter design to maximize the signal-to-interference-plus-noise-ratio (SINR) \cite{chen2009,Liu14,Cui14,Imani15}.
\par
This letter focuses on the last design approach for colocated MIMO radar. In this design framework, joint transmit and receive beamforming is investigated in \cite{Liu14} for an active array in the presence of signal-dependent interference. A sequential optimization algorithm is proposed to maximize the output SINR.
In \cite{Cui14}, joint transmit waveform and receive filter design is considered under the constant modulus and similarity constraint. Both works rely on exact knowledge of target and interferences. Indeed, the angle and INR of interferences can be obtained from knowledge-aided methods or estimated through previous scans of the space in high INR cases \cite{Aubry13,Duly13}. The known target angle assumption can be applied to the confirmation of an initial detection at some angle bin \cite{Antonio10}.
However, there are other situations where the target angle is unknown (e.g., weak target embedded in strong interferences), and the SINR should be averagely optimized over the uncertain area to avoid beampattern loss \cite{Tabrikian06}. Hence, angular-robust design must be considered and the robust design can also be used as an initial step for cognitive detection. In previous works \cite{Antonio10,Naghsh14}, robust waveform design has been studied for interpulse (or intrapulse) coding in radar by considering the unknown Doppler shift of target. Motivated by these works, we consider the problem of angular-robust design for colocated MIMO radar in the presence of signal-dependent interferences, which are induced by the interaction of transmit waveform with unwanted scatters. Based on the SINR criterion, transmit waveform and receive filter are jointly optimized to maximize the worst-case output SINR. Since the resulting problem is non-convex, cyclic optimization \cite{Stoica04} and semi-definite relaxation (SDR) \cite{Luo10} are used to solve it. Although the cyclic optimization converges to a locally optimal solution, it still can yield a good enough solution with higher worst-case SINR than the non-robust design, as illustrated in section IV. This is fundamentally different from the optimization problem in parameter estimation, in which the local convergence may significantly deteriorate the accuracy of estimation. SDR is a powerful approximation technique to solve a host of difficult non-convex problems with rank constraints. It is commonly used in radar signal processing problems, e.g., \cite{Aubry13,Duly13,Antonio10,Naghsh14}.
\par
\emph{Notations:} Matrices are denoted by bold capital letters, and vectors by bold lowercase letters. $(\cdot)^T$, $(\cdot)^c$ and $(\cdot)^H$ denote the transpose, conjugate and conjugate transpose, respectively. $\|\cdot\|$ denotes Euclidean norm. $\otimes$ denotes Kronecker product. $\bm{I}_L$ means
$L \times L$ identity matrix. $\mathbb{R}$ and $\mathbb{C}$ denotes the sets of all real numbers and complex numbers, respectively. $\delta(\cdot)$ represents Kronecker delta function. $\vec{\cdot}$ denotes vectorization operator. $\Re{\cdot}$ denotes the real part of the argument.
\section{Problem Formulation}
Consider a colocated MIMO radar system equipped with $N_T$ transmitters and $N_R$ receivers. Both the transmit and receive arrays are assumed to be uniform linear arrays with half-a-wavelength element-separation. Let $\S \in \C^{N_T \times N}$ denote the transmitted waveform matrix, where $N$ is the number of samples in the duration of the transmitted waveform.
For a particular range cell of interest, the received waveform matrix $\Y \!\in\! \C^{N_R \times N}$ from $N_R$ receivers is corrupted by $K$ signal-dependent interferences (e.g., other targets in a multi-target scenario \cite{Duly13}) from adjacent range cells with the additional noise, and is modeled as
\beq
   \Y = \alpha_0\ar{0}\at{0}^T\S + \sum_{k=1}^{K}\alpha_k\ar{k}\at{k}^T\S\J_{r_k} + \N \notag
\eeq
where
\begin{itemize}
\item
$\alpha_0$ and $\alpha_k$ are the complex amplitudes of the target and the $k$-th interference source, respectively.
\item
$\theta_0$ and $\theta_k$ are the direction of arrivals (DOA) of the target and the $k$-th interference source, respectively.
\item
$\ar{}\in \mathbb{C}^{N_R \times 1}$ is the receive steering vector defined by $\ar{} \triangleq \bgv{1, e^{j\pi\sin(\theta)},\cdots,e^{j\pi(N_R-1)\sin(\theta)}}^T$.
\item
$\at{}\in\mathbb{C}^{N_T \times 1}$ is the transmit steering vector defined by $\at{}\triangleq \bgv{1, e^{j\pi\sin(\theta)},\cdots,e^{j\pi(N_T-1)\sin(\theta)}}^T$.
\item
$\J_{r_k}, r_k \!\in\! \{-N+1, \cdots, -1, 0,1,\cdots,N-1\}$ is an $N$-by-$N$ shift matrix with $(l_1,l_2)$-th element $\J_{r}(l_1,l_2) \triangleq \delta(l_1 - l_2 - r)$. $r_k$ is the range cell index of the $k$-th interference source relative to the range cell of interest.
\item
$\N$ is spatially and temporally white circularly symmetric complex Gaussian noise with mean zero and variance $\sigma^2$.
\end{itemize}
\par
Let $\y = \vec{\Y}$, $\s = \vec{\S}$ and $\n = \vec{\N}$. The vectorization form of the measurement model is given by
    \equ{\y = \alpha_0\A(\theta_0)\s + \sum_{k=1}^{K}\alpha_k\B(\theta_k)\s + \n}
where $\A(\theta_0) = \I_N \otimes \big[\ar{0}\at{0}^T\big]$ and $\B(\theta_k) = \J_{r_k}^T \otimes \big[\ar{k}\at{k}^T\big]$. The SINR at the output of the receive filter $\w\in\mathbb{C}^{N_RN \times 1}$ is given by
\beq
    \obj(\s,\w,\theta_0) = \frac{\SNR \left|\w^H\A(\theta_0)\s\right|^2}{\w^H\bm{\Sigma}_I(\s)\w+\w^H\w} \label{objFunc}
\eeq
where $\bm{\Sigma}_I(\s) = \sum_{k=1}^{K} \INR_k\B(\theta_k)\s\s^H\B(\theta_k)^H$
with the signal-to-noise-ratio (SNR) of target and the interference-to-noise-ratio (INR) of $k$-th interference defined as $\SNR \triangleq \E{|\alpha_0|^2}/\sigma^2$ and $\INR_k \triangleq \E{|\alpha_k|^2}/\sigma^2$, respectively. \par
We assume that the angle and the INR of the interferences are all known or previously estimated, as with prior works \cite{Liu14,Cui14,Duly13}. We assume that the angle of target under test is known to lie in an angular sector $\Omega = [\theta_C-\Delta\theta, \theta_C+\Delta\theta]$ centred around $\theta_C$, where $\Delta\theta$ indicates the level of angular uncertainty. The goal is to maximize the worst-case SINR to improve the worst-case detection performance, under the waveform energy constraint $\|\s\|^2 = E$.
Therefore, the robust design of transmit waveform and receive filter can be formulated as the following max-min problem:
\beq
    \underset{\s, \w}{\max}\;\;\underset{\theta_0\in \Omega}{\min} \quad \obj(\s,\w,\theta_0) \quad \textrm{subject to}\quad \|\s\|^2 = E \label{mainprob1}
\eeq
Note that for the case of known target angle, (\ref{mainprob1}) reduces to the optimization problem in \cite{Liu14}.
\section{Max-Min Robust Design Algorithm}
In this section, we shall present our algorithm to solve the problem (\ref{mainprob1}).
To begin with, we make some mathematical transformations to the objective function of the optimization problem.
Define $\W \in \C^{N_R\times N}$ such that $\w = \vec{\W}$. Let $\p(\nu) = [\begin{matrix} 1,e^{j\nu},\cdots ,e^{j\nu(L-1)} \end{matrix}]^T$ with $L = N_R+N_T-1$ and $\nu = \pi\sin(\theta_0)$. Let $\H = [\begin{matrix} \tilde{\H}_1^T,\tilde{\H}_2^T,\cdots,\tilde{\H}_{N_R}^T\end{matrix}]^T$ where $\H \in \mathbb{R}^{N_RN_T \times L}$, $\tilde{\H}_k \in \mathbb{R}^{N_T \times L}, k=1,\cdots,N_R$, and the $(m,n)$-th element of $\tilde{\H}_k$ is defined by $\tilde{\H}_k(m,n) \triangleq \delta(n-m-k+1)$. Then, one can easily show that $\ar{0} \otimes \at{0} = \H\p(\nu)$. According to the property of Kronecker products that $\vec{\mb{C}\X\D} = (\D^T\otimes\mb{C})\vec{\X}$, we can show that
\begin{align}
    \w^H\A(\theta_0)\s &= \left(\A(\theta_0)^T\vec{\W^{\ast}}\right)^T\s \\
                      &= \mathrm{vec}\big(\at{0}\ar{0}^T\W^{\ast}\big)^T\s \\
                      &= \big((\W^H \otimes \I_{N_T})\mathrm{vec}\big(\at{0}\ar{0}^T\big)\big)^T\s \\
                      &= \s^T(\W^H \otimes \I_{N_T})\big(\ar{0} \otimes \at{0}\big) \\
                      %&= \big((\W^{\ast} \otimes \I_{N_T})\vec{\S}\big)^T\H\p(\nu)\\
                      & = \vec{\S\W^H}^T\H\p(\nu)
\end{align}
Then, it follows that $|\w^H\A(\theta_0)\s|^2 = \p(\nu)^H\G\p(\nu)$ where
\beq
\G \triangleq \H^H\vec{\S\W^H}^{\ast}\vec{\S\W^H}^T\H \label{G}
\eeq
Let $\S = \bgv{\s_1,\s_2, \cdots, \s_{N}}$ and $\W = \bgv{\w_1,\w_2,\cdots,\w_{N}}$. Using $\S\W^H = \sum_{n=1}^{N} \s_n\w_n^H$, we can write
\beq
    \vec{\S\W^H} = \sum_{n=1}^{N} \vec{\s_n\w_n^H} = \sum_{n=1}^{N}\w_n^{\ast}\otimes\s_n \label{SW}
\eeq
Define $\X = \s\s^H$ and $\V = \w\w^H$ with $\X \in \C^{N_TN \times N_TN}$ and $\V \in \C^{N_RN \times N_RN}$.  Partition $\X$ and $\V$ into a $N$-by-$N$ block matrix with $(n_1,n_2)$-th block denoted by $\X_{[n_1,n_2]} \in \C^{N_T\times N_T}$ and $\V_{[n_1,n_2]} \in \C^{N_R\times N_R}$, then it follows that
\beq
    \G(\X,\V) = \H^H \bigg(\sum_{1\leq n_1,n_2\leq N} \V_{[n_1,n_2]} \otimes \X_{[n_1,n_2]}^{\ast} \bigg)\H \label{GXV}
\eeq
where we use the notation $\G(\X,\V)$ to emphasize $\G$ as a function of $\X$ and $\V$. Moreover, using $\w^H\w = \tr{\V}\tr{\X}/E$ and $\w^H\bm{\Sigma}_I(\s)\w = \tr{\bm{\Sigma}_I(\V)\X}$ with $\bm{\Sigma}_I(\V) = \sum_{k=1}^{K}\INR_k \B(\theta_k)^H\V\B(\theta_k)$, it is easy to find that the denominator of (\ref{objFunc}) can be re-written as $\text{tr}\big(\big(\bm{\Sigma}_I(\V)+\frac{\tr{\V}}{E}\I_{N_TN}\big)\X\big)$.
Consequently, problem (\ref{mainprob1}) can be recast as
\beq
    \begin{cases}
        \;\underset{\X,\V}{\max}\;\;\underset{\nu \in \mathcal{I}}{\min}\quad &\dfrac{\p(\nu)^H\G(\X,\V)\p(\nu)}{\tr{\left(\bm{\Sigma}_I(\V)+\frac{\tr{\V}}{E}\I_{N_TN}\right)\X}} \\
        \;\textrm{subject to}\quad
        &\tr{\X} = E,\; \X \succeq \bm{0},\;\V \succeq \bm{0} \\
        & \rank{\X} = 1,\;\rank{\V} = 1
    \end{cases}
    \label{nonConvex}
\eeq
where $\mathcal{I} = [\nu_C-\Delta\nu, \nu_C+\Delta\nu]$ is the corresponding uncertain range of $\nu$ after parameter transformation. \par
\subsection{Optimization with respect to $\X$ and $\V$}
Since the rank constraint in (\ref{nonConvex}) is non-convex, we adopt the commonly-used SDR technique \cite{Luo10} to obtain a relaxed problem by dropping the rank-one constraint in (\ref{nonConvex}):
\beq
    \begin{cases}
        \;\underset{\X,\V}{\max}\;\;\underset{\nu \in \mathcal{I}}{\min}\quad &\dfrac{\p(\nu)^H\G(\X,\V)\p(\nu)}{\tr{\left(\bm{\Sigma}_I(\V)+\frac{\tr{\V}}{E}\I_{N_TN}\right)\X}} \\
        \;\textrm{subject to}\quad
        &\tr{\X} = E,\;\X \succeq \bm{0},\;\V \succeq \bm{0}
    \end{cases}
    \label{nonConvex2}
\eeq
or equivalently,
\beq
    \begin{cases}
        \underset{\U,\V,t,\gamma}{\max} \; &t \\
        \textrm{subject to}\; &\p(\nu)^H\G(\U,\V)\p(\nu) \geq t,\; \text{for}\;\;\forall \nu \in \mathcal{I} \\
        &\tr{\left(\bm{\Sigma}_I(\V)+\frac{\tr{\V}}{E}\I_{N_TN}\right)\U} = 1 \\
        &\tr{\U} = E\gamma,\; \gamma \ge 0 \\
        & \U \succeq \bm{0}\;\V \succeq \bm{0}
    \end{cases}
    \label{nonConvex3}
\eeq
where $\U = \gamma\X$. Let $\g = \bgv{g_0,g_1,\cdots,g_{L-1}}^T$ with $g_l = \sum_{k=1}^{N-l}\G(\U,\V)_{l+k,k},\,l=0,1,\cdots,L\!-\!1$. One can also show that the constraint $\p(\nu)^H\G\p(\nu) \geq t$ in (\ref{nonConvex3}) is equivalent to
\beq
    f(\nu) = g_0 - t + 2\text{Re}\bigg\{\sum_{l=1}^{L-1} g_le^{-jl\nu}\bigg\} \geq 0 \label{polynomial}
\eeq
The optimization problem (\ref{nonConvex3}) is still non-convex and it includes infinitely many quadratic constraints as $\nu \in \mathcal{I}$.
To deal with this problem, we resort to an equivalent semi-definite programming (SDP) representation for the nonnegativity constraint of the trigonometric polynomial in (\ref{polynomial}) based on \cite[Theorem 3.4]{roh2006}, which is quoted below as a lemma.
\begin{lemma}
    The trigonometric polynomial $\tilde{f}(\omega) = h_0 + 2\mathrm{Re}\big\{\sum_{l=1}^{L-1} h_le^{-j\omega l}\big\}$ is non-negative over $[\alpha-\beta,\alpha+\beta]$ (with $0< \beta <\pi$) iff there exists an $L \times L$ Hermitian matrix $\Z_1 \succeq 0$ and an $(L-1) \times (L-1)$ Hermitian matrix $\Z_2 \succeq 0$ such that
    \beq
        \h = \F_1^H\left( \diag{\F_1\Z_1\F_1^H} +\d \odot \diag{\F_2\Z_2\F_2^H} \right)
    \eeq
    where $\h = \bgv{h_0,h_1,\cdots,h_{L-1}}^T$, $\d = \bgv{d_0,d_1,\cdots,d_{Q-1}}^T$ with $d_q = \cos(2\pi q/Q-\alpha) -\cos(\beta)$, $\F_1 = \bgv{\f_0,\f_1,\cdots,\f_{L-1}}$ and $\F_2 = \bgv{\f_0,\f_1,\cdots,\f_{L-2}}$ where $\f_l = \bgv{1,e^{-j2\pi l/Q},\cdots,e^{-j2\pi l(Q-1)/Q}}^T$ with $Q \geq 2L-1$.\label{lemma1}
\end{lemma}
\par
Based on Lemma \ref{lemma1}, cyclic optimization \cite{Stoica04} can then be performed to tackle problem (\ref{nonConvex3}) iteratively. To be specific, we perform the optimization with respect to $\U$ for some fixed $\V$, and then conduct it with respect to $\V$ for fixed $\U$, repeatedly. To this end, let $\alpha = \nu_C$, $\beta = \Delta\nu$ and $\h = \g - t\e_1$ in Lemma \ref{lemma1}, where $\e_1$ is an $L\times1$ vector with the first component being one and the others zero.
For fixed $\V$, the optimization with respect to $\U$ for (\ref{nonConvex3}) can be represented by the following SDP:
%\textcolor{red}{which can be solved by the CVX toolbox \cite{cvx}}:\par
\beq
    \begin{cases}
        \;\underset{\U,\Z_1,\Z_2,t}{\max}  &t \\
        \;\textrm{subject to} &\g - t\e_1 = \F_1^H\left( \diag{\F_1\Z_1\F_1^H} \right. \\
        &\qquad\qquad\:\: + \left. \d \odot \diag{\F_2\Z_2\F_2^H} \right) \\
        &\tr{\Big(\bm{\Sigma}_I(\V)+\frac{\tr{\V}}{E}\I_{N_TN}\Big)\U} = 1 \\
        & \U \succeq \bm{0},\; \Z_1 \succeq \bm{0},\; \Z_2 \succeq \bm{0}
    \end{cases}
    \label{Uoptm}
\eeq
Let $\U^{\star}$ denote the optimal solution of $\U$ to (\ref{Uoptm}). Then, the optimal solution of $\X$ is equal to
$E\U^{\star}/\tr{\U^{\star}}$. The SDP problem can be solved efficiently using the interior point methods in polynomial time \cite{boyd2004convex}. In the simulations, we use the MATLAB toolbox CVX \cite{cvx} to solve problem (\ref{Uoptm}).\par
Since the denominator of the objective function in (\ref{nonConvex2}) can also be expressed as $\tr{\left(\bm{\Sigma}_I(\X)+\I_{N_RN}\right)\V}$ with $\bm{\Sigma}_I(\X) = \sum_{k=1}^{K}\INR_k \B(\theta_k)\X\B(\theta_k)^H$, the optimization with respect to $\V$ for fixed $\X$ can be cast as a similar SDP as below.
\beq
    \begin{cases}
        \;\underset{\V,\Z_1,\Z_2,t}{\max} \quad &t \\
        \;\textrm{subject to}\quad &\g - t\e_1 = \F_1^H\left( \diag{\F_1\Z_1\F_1^H} \right. \\
        &\qquad\qquad\:\: + \left. \d \odot \diag{\F_2\Z_2\F_2^H} \right) \\
        &\tr{\left(\bm{\Sigma}_I(\X)+\I_{N_RN}\right)\V} = 1 \\
        & \V \succeq \bm{0},\; \Z_1 \succeq \bm{0},\; \Z_2 \succeq \bm{0}
    \end{cases}
    \label{Voptm}
\eeq
\par
By starting from a random initial point and cyclically solving (\ref{Uoptm}) and (\ref{Voptm}) until the SINR improvement is negligible, the objective function value is non-decreasing and the convergence of the algorithm can be guaranteed \cite{Naghsh14}.
The cyclic optimization converges to a point which is not only the local optimum, but also the global optimum along the $\X$ dimension and the $\V$ dimension separately \cite{chen2009}.
To obtain a more accurate result, one can perform this procedure with a large number of random initializations and then select the best $(\X,\V)$. In section IV, numerical examples show that the proposed algorithm is insensitive to initial values.
\par
\subsection{Synthesis of $\s$ and $\w$ from $\X$ and $\V$} \label{sysAlg}
Let $(\X^{\star}, \V^{\star})$ denote the solution of (\ref{nonConvex2}) using the cyclic optimization.
If both $\X^{\star}$ and $\V^{\star}$ are rank-one,
the transmit waveform $\s^{\star}$ and receive filter $\w^{\star}$
can be obtained by the eigen-decomposition of $\X^{\star} = \s^{\star}(\s^{\star})^H$ and $\V^{\star} = \w^{\star}(\w^{\star})^H$. In this case, the rank-one relaxation in (\ref{nonConvex3}) is tight and the solution is optimal. Otherwise, a suboptimal procedure can be adopted following a recently proposed algorithm in \cite{Naghsh14}.
The basic idea of the algorithm is based on the fact that $\obj(\s,\w)$ is a scaled version of the numerator $\p(\nu)^H\G(\X,\V)\p(\nu)$, or $\tr{\X\A(\theta_0)^H\V\A(\theta_0)}$ equivalently. Then
$\s^{\star}$ and $\w^{\star}$ should be designed to let $|(\w^{\star})^H\A(\theta_0)\s^{\star}|^2$ well approximate the shape of $\tr{\X^{\star}\A(\theta_0)^H\V^{\star}\A(\theta_0)}$, while imposing constraint on the denominator.
Interested readers can refer to \cite{Naghsh14} for detailed motivation. To make the letter self-contained, we shall present the synthesis algorithm for our problem in the sequel.
\par
Consider the value of $\tr{\X^{\star}\A(\theta_0)^H\V^{\star}\A(\theta_0)}$ evaluated on DOAs $\{\vartheta_1,\vartheta_2,\cdots,\vartheta_M\}$ ``uniformly distributed'' on $\Omega$:
\beq
    c_m = \tr{\X^{\star}\A(\vartheta_m)^H\V^{\star}\A(\vartheta_m)},\; m = 1,2,\cdots, M
\eeq
Let $\T_m = \A(\vartheta_m)^H\V^{\star}\A(\vartheta_m)$, $\Q_m\Q_m^H = \T_m$ and define $M$ auxiliary unit-norm vectors $\q_1,\q_2,\cdots,\q_M$.
Then, the synthesis of $\s$ can be formulated as
\beq
    \begin{cases}
       \underset{\bar{\s},\,\q_1,\cdots,\q_M}{\min} &\sum_{m=1}^M\|\Q_m\bar{\s} - \sqrt{c_m} \q_m\|^2 \\
       \;\textrm{subject to} &\bar{\s}^H\left(\bm{\Sigma}_I(\V^{\star})+\frac{\tr{\V^{\star}}}{E}\I_{N_TN}\right)\bar{\s} \leq \zeta^{\star} \\
       &\|\q_m\| = 1,\; 1 \leq m \leq M
    \end{cases} \label{sysS}
\eeq
where $\zeta^{\star} \triangleq \text{tr}\big(\big(\bm{\Sigma}_I(\V^{\star})+\frac{\tr{\V^{\star}}}{E}\I_{N_TN}\big)\X^{\star}\big)$. This problem can be solved using cyclic minimization.
For a fixed $\bar{\s}$, the solution to (\ref{sysS}) is given by $\q_m = \frac{\Q_m\bar{\s}}{\|\Q_m\bar{\s}\|}, m = 1,\cdots,M$. For fixed $\q_m, m = 1,\cdots,M$, problem (\ref{sysS}) reduces to a quadratically constrained quadratic program (QCQP) that can be solved by the CVX package \cite{cvx}. The initial value of $\bar{\s}$ can be chosen as the eigenvector of $\X^{\star}$ corresponding to the largest eigenvalue. Let $\bar{s}^{\star}$ denote the optimal solution to (\ref{sysS}), the optimal transmit waveform $\s^{\star}$ is given by
$\s^{\star} = \sqrt{E}\bar{s}^{\star}/{\|\bar{s}^{\star}\|}$,
considering the energy constraint on $\s$. \par
Analogously, let $\widetilde{\T}_m = \A(\vartheta_m)\X^{\star}\A(\vartheta_m)^H$, $\widetilde{\Q}_m\widetilde{\Q}_m^H = \widetilde{\T}_m$, and $\eta^{\star} \triangleq \tr{\left(\bm{\Sigma}_I(\X^{\star})+\I_{N_RN}\right)\V^{\star}}$,
the synthesis of $\w$ is similar to problem (\ref{sysS}):
\beq
    \begin{cases}
       \underset{\w,\,\widetilde{\q}_1,\cdots,\widetilde{\q}_M}{\min} \;\;&\sum_{m=1}^M \|\widetilde{\Q}_m\w - \sqrt{c_m} \widetilde{\q}_m \|^2 \\
       \;\textrm{subject to} \;\; &\w^H\left(\bm{\Sigma}_I(\X^{\star})+\I_{N_RN}\right)\w \leq \eta^{\star} \\
       &\|\widetilde{\q}_m\| = 1,\; 1 \leq m \leq M
    \end{cases}
    \label{sysW}
\eeq
which can also be solved using the cyclic minimization.
\par
We also note that the randomized method \cite{Luo10} can also be used to obtain an approximate $\s^{\star}$ and $\w^{\star}$ in the non-rank-one case. Similar applications can be found in
\cite{Antonio09,Cui14,Karbasi15}. The synthesis algorithm based on the randomized method for our problem is shown in Algorithm \ref{alg}. \par

\begin{algorithm}[!htb]
\caption{Synthesis algorithm based on randomized method}
\label{alg}
\begin{algorithmic}[1]
\REQUIRE ~$\X^{\star}$ and $\V^{\star}$
\ENSURE ~A randomized approximate solution $\s^{\star}$ and $\w^{\star}$
\IF{$\text{rank}(\V^{\star}) = 1$}
\STATE{find $\w^{\star}$ via eigen-decomposition $\V^{\star} = \w^{\star}(\w^{\star})^H$}
\ELSE
\STATE{draw $R$ random vectors $\w_j$ from the complex Gaussian distribution $\mathcal{CN}(\bm{0},\V^{\star}), j = 1,2,\cdots,R$}
\STATE{calculate \[\xi_j = \underset{\theta_0\in \Omega}{\min}\;\frac{ \w_j^H\A(\theta_0)\X^{\star}\A(\theta_0)^H\w_j}{\w_j^H\bm{\Sigma}_I(\X^{\star})\w_j+\w_j^H\w_j},\;j = 1,\cdots,R\]  where $\bm{\Sigma}_I(\X^{\star}) = \sum_{k=1}^{K}\INR_k \B(\theta_k)\X^{\star}\B(\theta_k)^H$.}
\STATE{let $\w^{\star} = \w_{j_{\max}}$ where \[ j_{\max} = \arg\,\underset{1 \le j \le R}{\max}\; \xi_j.\]}
\ENDIF
\IF{$\text{rank}(\X^{\star}) = 1$}
\STATE{find $\s^{\star}$ via eigen-decomposition $\X^{\star} = \s^{\star}(\s^{\star})^H$}
\ELSE
\STATE{draw $R$ random vectors $\s_i$ from the complex Gaussian distribution $\mathcal{CN}(\bm{0},\X^{\star}), i = 1,2,\cdots,R$}
\STATE{calculate $\bar{\s}_i = \frac{\sqrt{E}\s_i}{\|\s_i\|}$ and \[
\zeta_i = \underset{\theta_0\in \Omega}{\min}\;\frac{ |(\w^{\star})^H\A(\theta_0)\bar{\s}_i|^2}{(\w^{\star})^H\bm{\Sigma}_I(\bar{\s}_i)\w^{\star}+(\w^{\star})^H\w^{\star}}, \;i = 1,\cdots,R\] where $\bm{\Sigma}_I(\bar{\s}_i) = \sum_{k=1}^{K} \INR_k\B(\theta_k)\bar{\s}_i\bar{\s}_i^H\B(\theta_k)^H$.}
\STATE{let $\s^{\star} = \bar{\s}_{i_{\max}}$ where
\[i_{\max} = \arg\,\underset{1 \le i \le R}{\max}\; \zeta_i.\]}
\ENDIF
\end{algorithmic}
\end{algorithm}
Prior results on the tightness of SDR
\cite{HuangPalomar10,Luo10} show that for a separable SDP \cite[eq. (28)]{Luo10} with $P$ semi-definite variables and $J$ constraints, there exists a rank-one optimal solution if $J \le P + 2$. But this can not guarantee the existence of rank-one solution for our problem, since $P = 3$ and $J = 2L-1$ for problem (\ref{Uoptm}) and (\ref{Voptm}) in the form of \cite[eq. (28)]{Luo10}. Nevertheless, we emphasize that as with in \cite{Naghsh14}, one can empirically observe that both $\X^{\star}$ and $\V^{\star}$ are rank-one for most of the random initializations as along as $\Omega \cap \Omega_c = \emptyset$, where $\Omega_c$ denotes the set of all interferences angles. \par
\section{Numerical Examples}
In this section, numerical examples are conducted to examine the performance of the proposed method. In all examples, we assume that $30$ interferences are present with the range and angle pair $(r_k,\theta_k)$ generated from all possible combinations of $\{-2,-1,0,1,2\} \times \{-60^{\circ}, -50^{\circ}, -40^{\circ}, 40^{\circ}, 60^{\circ}, 70^{\circ}\}$. The INR of all interferences is $30$ dB.
\par
\begin{figure}[!htb]
\centering
\subfigure[]{
\begin{minipage}[b]{0.2\textwidth}
\includegraphics[width=1\textwidth]{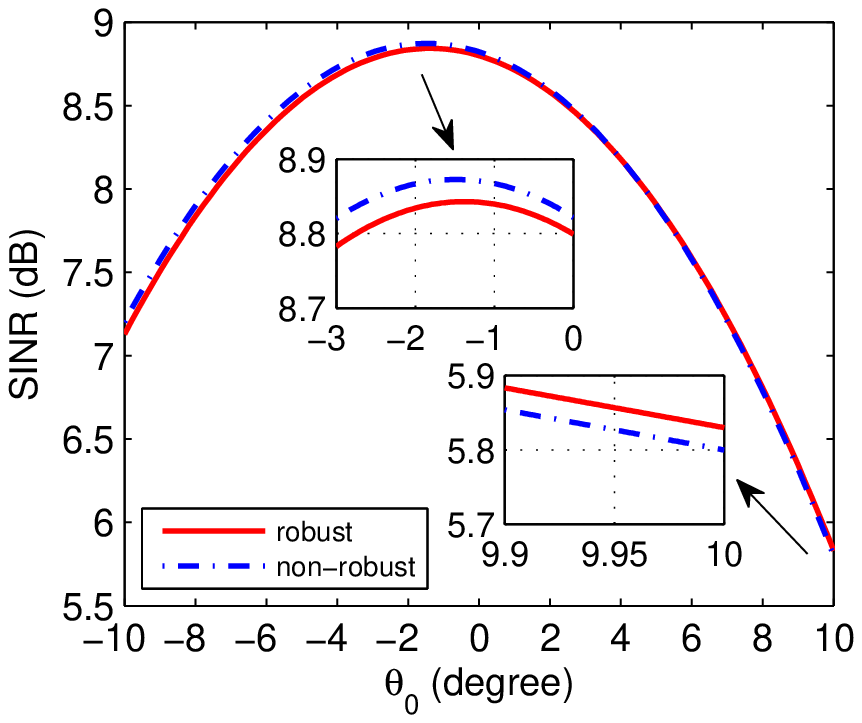}
\end{minipage}
\label{fig1a}
}
\centering
\subfigure[]{
\begin{minipage}[b]{0.2\textwidth}
\includegraphics[width=1\textwidth]{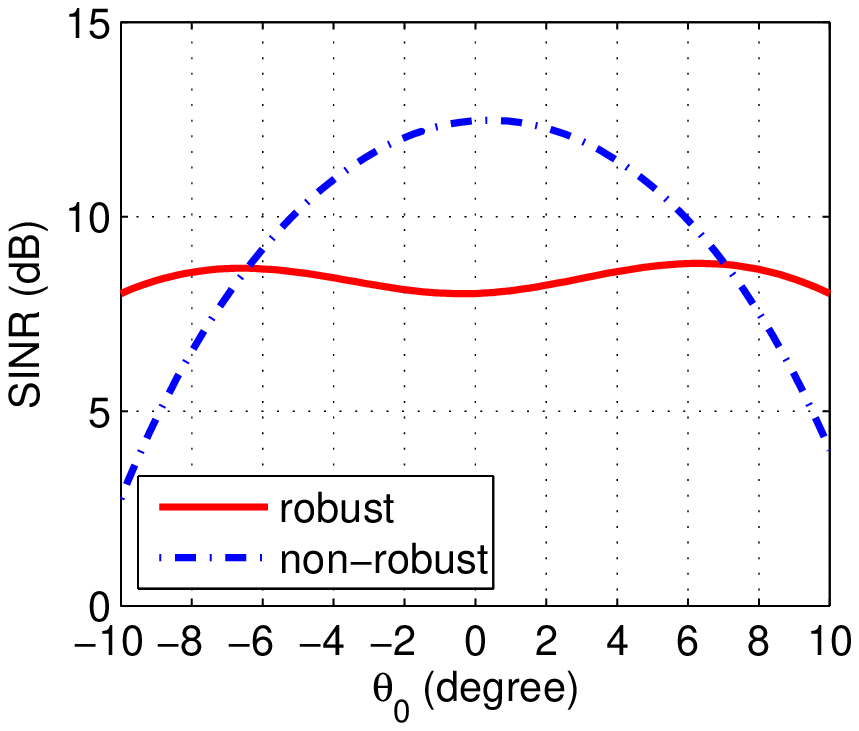}
\end{minipage}
\label{fig1b}
}
\centering
\subfigure[]{
\begin{minipage}[b]{0.2\textwidth}
\includegraphics[width=1\textwidth]{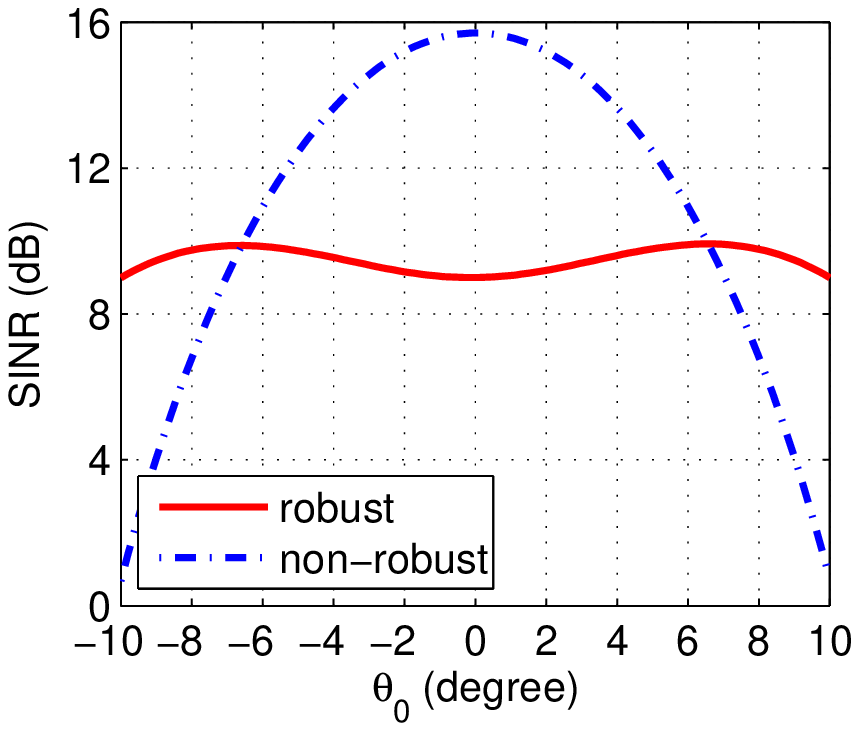}
\end{minipage}
\label{fig1c}
}
\centering
\subfigure[]{
\begin{minipage}[b]{0.2\textwidth}
\includegraphics[width=1\textwidth]{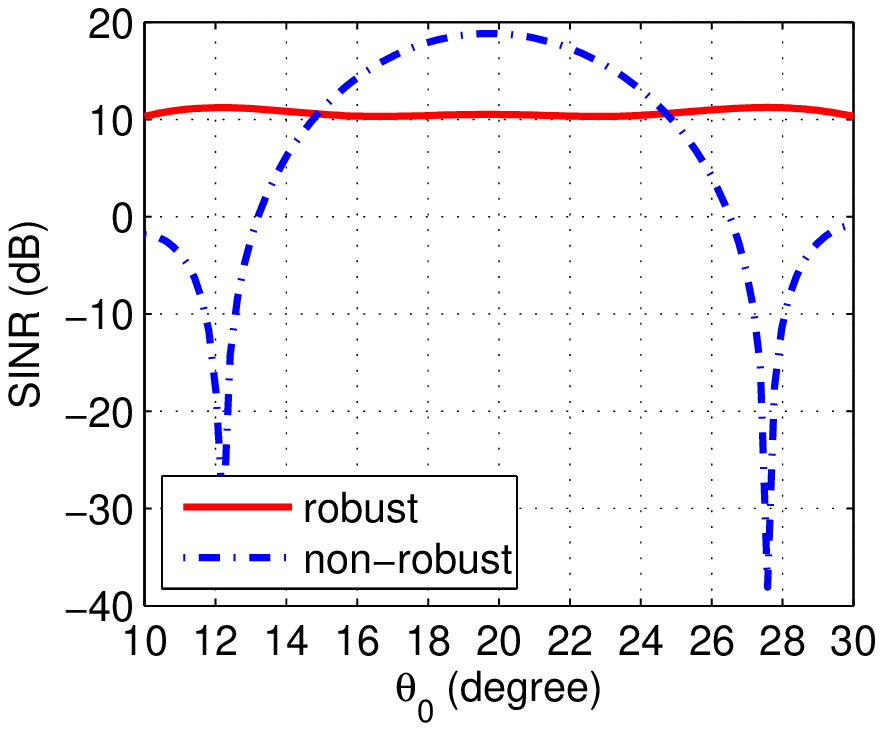}
\end{minipage}
\label{fig1d}
}
\caption{Comparisons of the output SINR. $\Delta\theta = 10^{\circ}$. $E = N = 20$. $\SNR = -15 \text{dB}$. (a) $N_R = N_T = 4$, $\theta_C = 0^{\circ}$; (b) $N_R = 4$, $N_T = 8$, $\theta_C = 0^{\circ}$; (c) $N_R = N_T = 8$, $\theta_C = 0^{\circ}$; (d) $N_R = 8, N_T = 16$, $\theta_C = 20^{\circ}$.} \label{fig1}
\end{figure}
\begin{figure}[!htb]
\centering
\subfigure{
\begin{minipage}[b]{0.2\textwidth}
\includegraphics[width=1\textwidth]{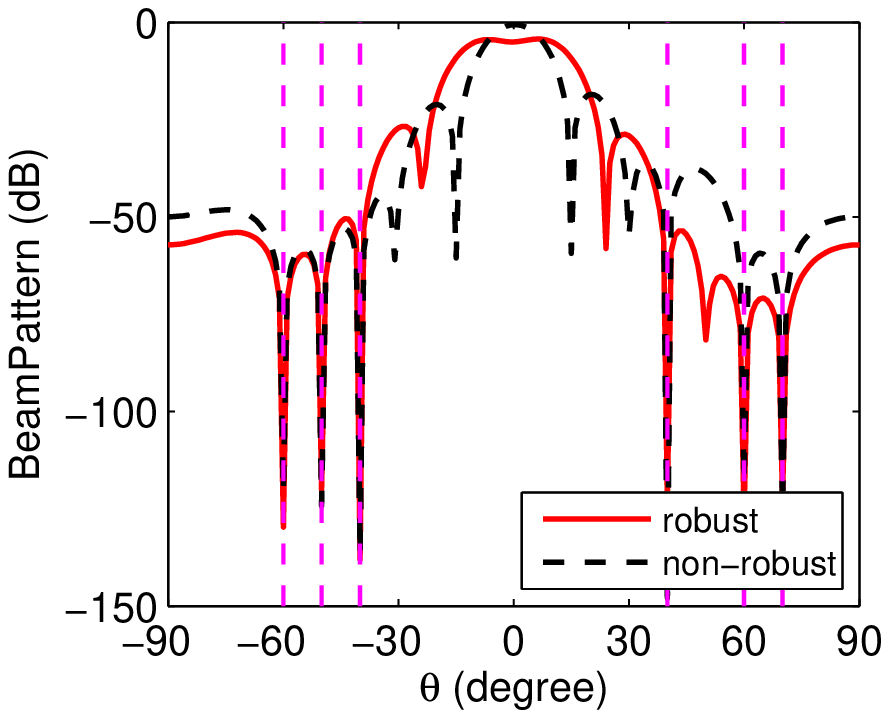}
\end{minipage}
}
\subfigure{
\begin{minipage}[b]{0.2\textwidth}
\includegraphics[width=1\textwidth]{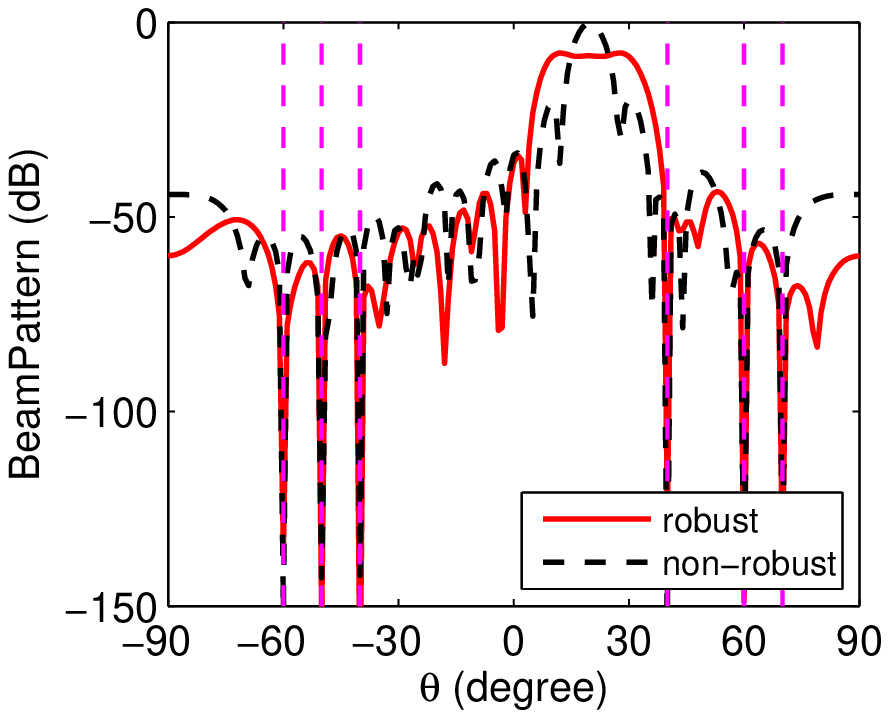}
\end{minipage}
}
\caption{Comparisons of the beampattern $P(\theta)$. (left) $N_R = N_T = 8$, $\theta_C = 0^{\circ}$; (right) $N_R = 8$, $N_T = 16$, $\theta_C = 20^{\circ}$.} \label{fig2}
\end{figure}

In Fig. \ref{fig1}, the output SINR as a function of $\theta_0$ for the non-robust design and the proposed robust design are compared under four different parameters. For the non-robust design, the assumed a-prior target angle is set to be $\theta_C$ and the optimization algorithm is based on the method presented in \cite{Liu14}.
It is shown that the robust design improves the worst-case SINR performance significantly at the cost of peak-SINR degradation. For fixed $\Delta\theta$ and $N$, the superiority of robust design increases with the number of transmitters or receivers.
In Fig. \ref{fig2}, we depict the beampattern $P(\theta) = \frac{\|\w^H\A(\theta)\s\|^2}{N_RN_T\|\w\|^2\|\s\|^2}$ for parameter settings in Fig. \ref{fig1c} and Fig. \ref{fig1d} as an example. One can observe that both robust and non-robust design can produce nulls near the DOAs of interferences. From Fig. \ref{fig1} and Fig. \ref{fig2}, we see that when $\Delta\theta$ is large enough relative to the beamwidth, the robust design can form a wide and flat beam over the uncertain space area to bring robustness. Both $\X^{\star}$ and $\V^{\star}$ are rank-one in this example.
\par

\begin{figure}[!htb]
\centering
\subfigure{
\begin{minipage}[b]{0.2\textwidth}
\includegraphics[width=1\textwidth]{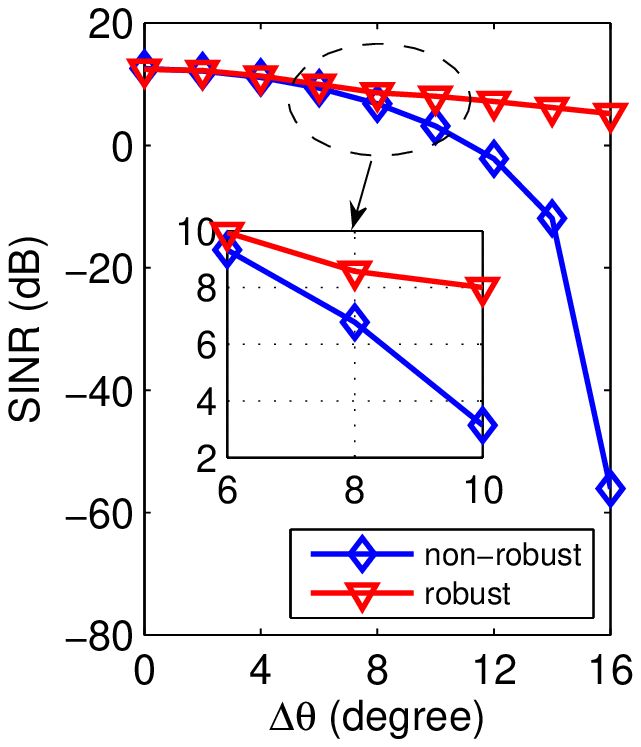}
\end{minipage}
}
\subfigure{
\begin{minipage}[b]{0.2\textwidth}
\includegraphics[width=1\textwidth]{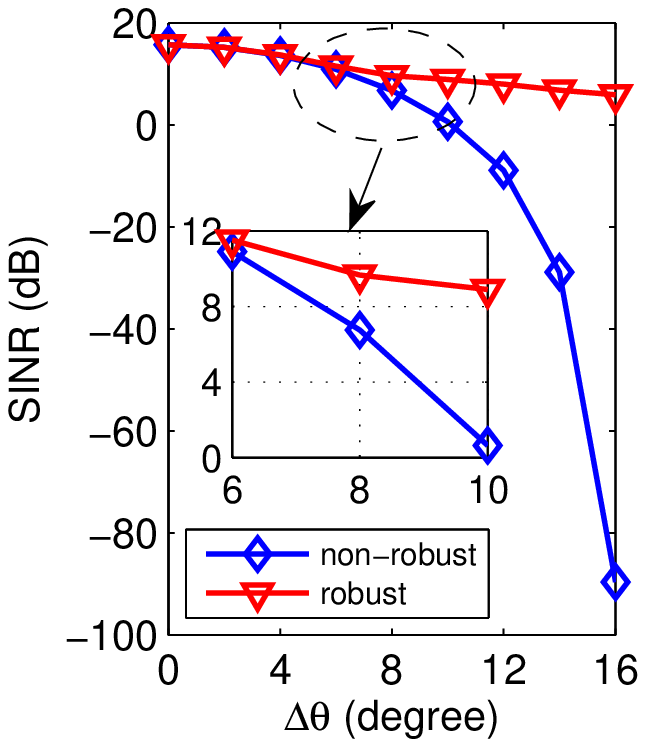}
\end{minipage}
}
\vspace{-3pt}
\caption{The worst-case output SINR versus the angle uncertainty. $\theta_C = 0^{\circ}$. $E = N = 20$. $\SNR = -15 \text{dB}$. (left) $N_R = 4$, $N_T = 8$; (right) $N_R = N_T = 8$. } \label{fig3}
\end{figure}

In Fig. \ref{fig3}, we plot the worst-case SINR versus the target angle uncertainty $\Delta\theta$. As expected, a wider range of target angle uncertainty leads to a worse SINR. The impact of $\Delta\theta$ on the worst-case SINR performance of non-robust design is more prominent, which suffers a sharp decline as $\Delta\theta$ increases. This is due to the effect of the first null near the main lobe. In this example, both $\X^{\star}$ and $\V^{\star}$ are rank-one.

\begin{figure}[!htb]
\centering
\subfigure[]{
\begin{minipage}[b]{0.2\textwidth}
\includegraphics[width=1\textwidth]{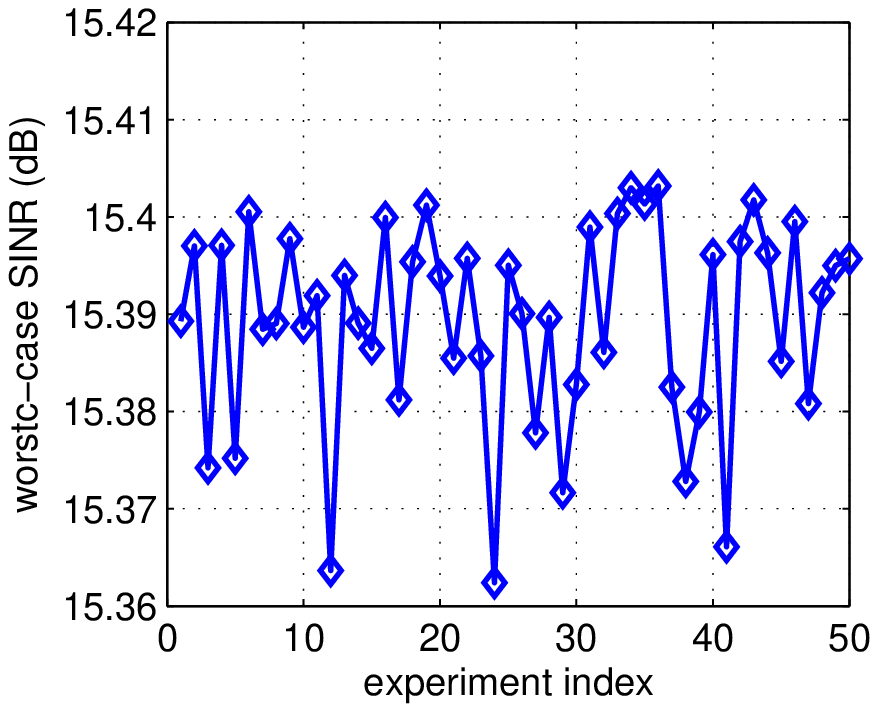}
\end{minipage}
\label{fig4a}
}
\centering
\subfigure[]{
\begin{minipage}[b]{0.2\textwidth}
\includegraphics[width=1\textwidth]{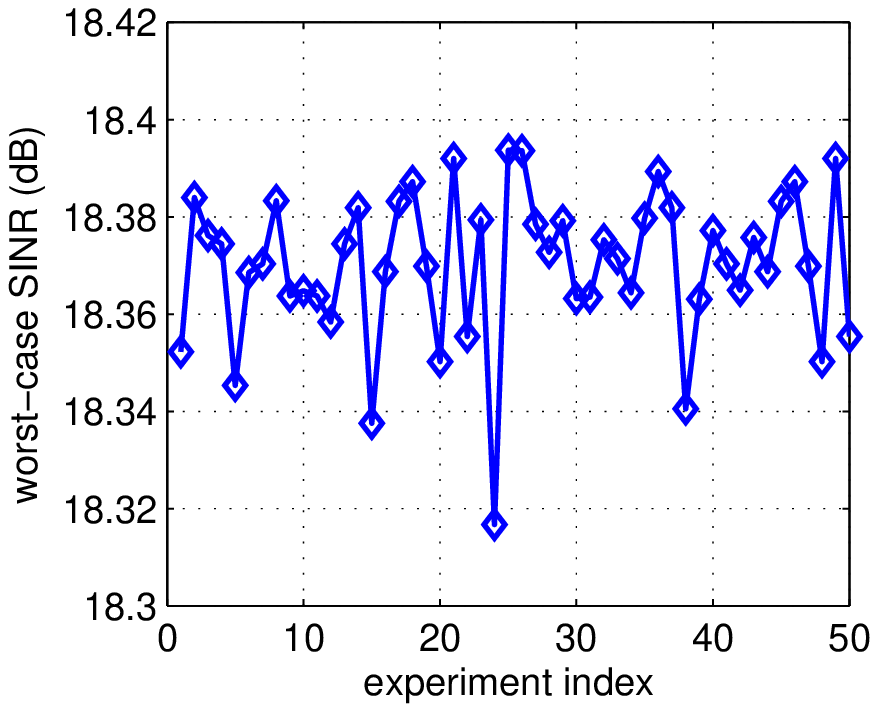}
\end{minipage}
\label{fig4b}
}
\centering
\subfigure[]{
\begin{minipage}[b]{0.2\textwidth}
\includegraphics[width=1\textwidth]{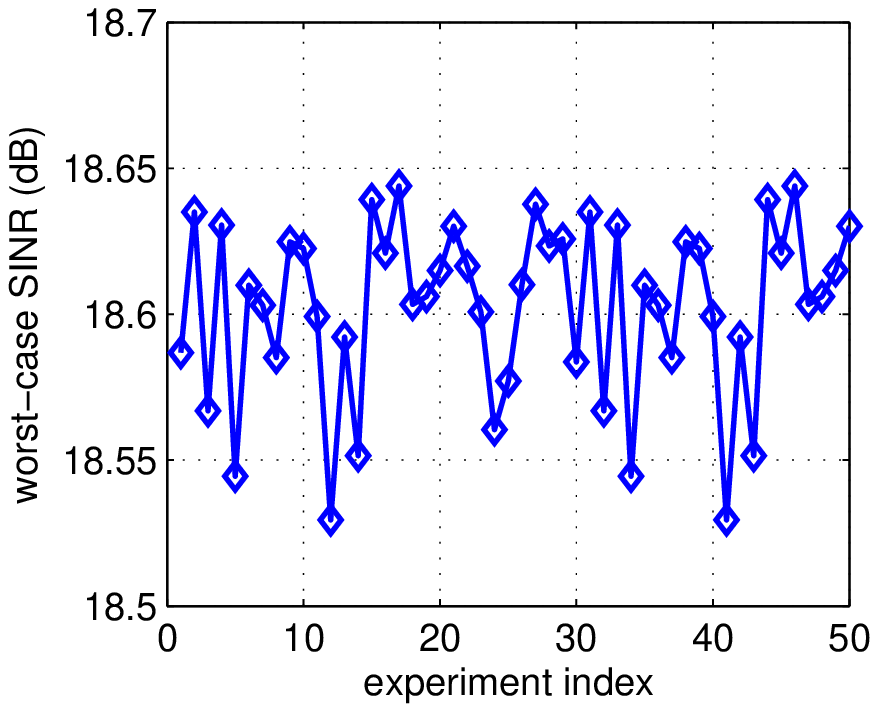}
\end{minipage}
\label{fig4c}
}
\centering
\subfigure[]{
\begin{minipage}[b]{0.2\textwidth}
\includegraphics[width=1\textwidth]{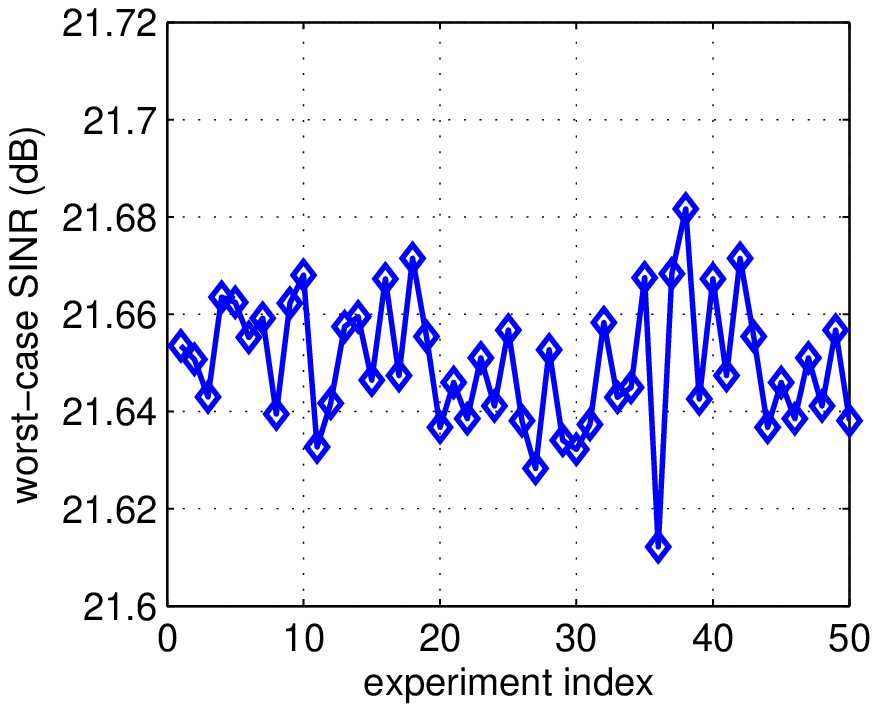}
\end{minipage}
\label{fig4d}
}
\caption{The effect of initial values on cyclic optimization. $E = N_TN$, $\SNR = 1/L$, $\theta_C = 0^{\circ}$ and $\Delta\theta = 10^{\circ}$. (a) $N_R = N_T = 4$, $N = 10$; (b) $N_R = N_T = 4$, $N = 20$; (c) $N_R = 4$, $N_T = 8$, $N = 10$. (d) $N_R = 4$, $N_T = 8$, $N = 20$.} \label{fig4}
\end{figure}

In Fig. \ref{fig4}, we investigate the effect of initial values on the cyclic optimization of $\X$ and $\V$. We plot the worst-case SINR for the relaxed problem (\ref{nonConvex3}) under $\Gamma = 50$ different random initializations.
Four different parameter settings are considered.
The cyclic optimization is stopped if either the increment of the worst-case SINR between two iterations is less than $5\times10^{-3}$ or the maximum number of iterations reaches. The maximum number of iterations of the cyclic optimization is set to $150$. We can see that the worst-case SINRs under different initializations are very close.
Let $\mathcal{T} = \{t^{(1)}, t^{(2)}, \cdots, t^{(\Gamma)}\}$ denote the worst-case SINRs from $\Gamma$ random initializations. We define the following metric
\beq
    \mathcal{L} \triangleq \frac{\max(\mathcal{T}) - \min(\mathcal{T})}{\text{mean}(\mathcal{T})}
\eeq
to evaluate the variation of $\mathcal{T}$, where $\max(\mathcal{T})$, $\min(\mathcal{T})$ and $\text{mean}(\mathcal{T})$ denote the maximum, minimum and mean value of $\mathcal{T}$, respectively. The values of $\mathcal{L}$ for the four cases are equal to $0.016$, $0.0177$, $0.0263$ and $0.016$, respectively. One can see that in our problem, the cyclic optimization is quite insensitive to the initialization.
\par

\begin{figure}[!htb]
\centering
\centerline{\includegraphics[width=5.8cm]{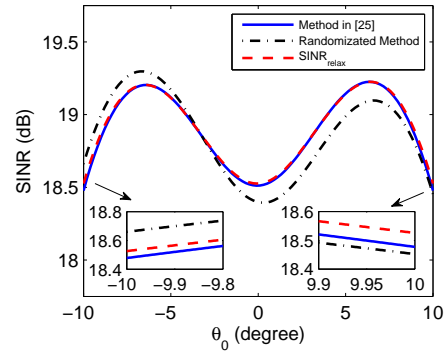}}
\caption{The SINR performance of the synthesis algorithm.} \label{fig5}
\vspace{-3pt}
\end{figure}

In Fig. \ref{fig5}, we illustrate the performance of the synthesis algorithm in the non-rank-one case, which seldom happens in our experiments. In this example, the parameter settings are the same as in Fig. \ref{fig4c}.
Under a certain random initialization, the cyclic optimization provides a solution with $\text{rank}(\X^{\star}) = 2$ and $\text{rank}(\V^{\star}) = 1$. The receive filter $\w^{\star}$ is obtained based on eigen-decomposition, and the transmit waveform $\s^{\star}$ is obtained via the synthesis algorithm. The performance of synthesis algorithm based on solving problem (\ref{sysS}) (denoted Method 1) and the algorithm based on randomized method (denoted Method 2) are compared. We plot their corresponding SINRs as a function of $\theta_0$ according to (\ref{objFunc}).
For the Method 1, the number of DOA samples $M$ is set to $41$ and the number of iterations to solve (\ref{sysS}) is $50$.
For the Method 2, the number of random samples is set to be $1000$.
We also plot the $\text{SINR}_{\text{relax}}(\theta_0) \triangleq \frac{\SNR\ \tr{\X^{\star}\A(\theta_0)^H\V^{\star}\A(\theta_0)}}{\tr{\left(\bm{\Sigma}_I(\X^{\star})+\I_{N_RN}\right)\V^{\star}}}$ as a benchmark for comparison. We can observe that their SINR performance are very close, and both synthesis algorithms yield a good solution in the non-rank-one case. We can also see that the SINR curve of Method 1 matches well with $\text{SINR}_{\text{relax}}(\theta_0)$. The worst-case SINRs for Method 1, Method 2 and $\text{SINR}_{\text{relax}}(\theta_0)$ are $18.478\,\text{dB}$, $18.394\,\text{dB}$ $18.526\,\text{dB}$, respectively. \par

\section{Conclusions}
A method for angular-robust joint design of transmit waveform and receive filter is proposed to maximize the worst-case SINR performance. The proposed method exhibits a considerable performance increment over the non-robust design via numerical examples. Future work will concentrate on the robust design with respect to the interferences uncertainty.

\bibliographystyle{IEEEtran}
\bibliography{refs}

\end{document}